\documentclass[a4paper,twoside]{article}

\usepackage{epsfig}
\usepackage[outdir=./]{epstopdf}
\usepackage{subcaption}
\usepackage{calc}
\usepackage{amssymb}
\usepackage{amstext}
\usepackage{amsmath}
\DeclareMathOperator*{\argminA}{arg\,min} 
\usepackage{amsthm}
\usepackage{multicol}
\usepackage{pslatex}
\usepackage{apalike}
\usepackage{SCITEPRESS}     

\begin{document}

\title{Restoring Eye Contact to the Virtual Classroom with Machine Learning}

\author{\authorname{Ross Greer\sup{1}, Shlomo Dubnov\sup{1}}
\affiliation{\sup{1}Department of Electrical \& Computer Engineering, University of California, San Diego, USA}
\affiliation{\sup{2}Department of Music, University of California, San Diego, USA}
\email{\{regreer, sdubnov\}@ucsd.edu}
}


\abstract{Nonverbal communication, in particular eye contact, is a critical element of the music classroom, shown to keep students on task, coordinate musical flow, and communicate improvisational ideas. Unfortunately, this nonverbal aspect to performance and pedagogy is lost in the virtual classroom. In this paper, we propose a machine learning system which uses single instance, single camera image frames as input to estimate the gaze target of a user seated in front of their computer, augmenting the user's video feed with a display of the estimated gaze target and thereby restoring nonverbal communication of directed gaze. The proposed estimation system consists of modular machine learning blocks, leading to a target-oriented (rather than coordinate-oriented) gaze prediction. We instantiate one such example of the complete system to run a pilot study in a virtual music classroom over Zoom software. Inference time and accuracy meet benchmarks for videoconferencing applications, and quantitative and qualitative results of pilot experiments include improved success of cue interpretation and student-reported formation of collaborative, communicative relationships between conductor and musician.}

\onecolumn \maketitle \normalsize \setcounter{footnote}{0} \vfill

\section{\uppercase{Introduction}}
\label{sec:introduction}

Teaching music in virtual classrooms for distance learning presents a variety of challenges, the most readily apparent being the difficulty of high rate technical synchronization of high quality audio and video to facilitate interactions between teachers and peers \cite{9238238}. Yet, often left out of this discussion is the richness of nonverbal communication that is lost over teleconferencing software, an element which is crucial to musical environments. Conductors provide physical cues, gestures, and facial expressions, all of which are directed by means of eye contact; string players receive eye contact and in response make use of peripheral and directed vision to anticipate, communicate, and synchronize bow placement; wind and brass instrumentalists, as well as singers, receive conducted cues to coordinate breathe and articulation. While the problem of \textit{technical synchronization} is constrained by the provisions of internet service providers and processing speeds of conferencing software servers and clients, the deprivation of nonverbal communication for \textit{semantic synchronization}, that is, coordination of musicians within the flow of ensemble performance, can be addressed modularly to restore some pedagogical and musical techniques to distance learning environments. 

This problem is not limited to the musical classroom; computer applications which benefit from understanding human attention, such as teleconferencing, advertising, and driver-assistance systems, seek to answer the question ‘Where is this person looking?’ \cite{frischen2007gaze}. Eye contact is a powerful non-verbal communication tool that is lost in video calls, because a user cannot simultaneously meet the eyes of another user as well as their own camera, nor can a participant ascertain if they are the target of a speaker's gaze. Having the ability to perceive and communicate intended gaze restores a powerful, informative communication channel to inter-personal interactions during live distance learning. 

The general problem of third-person gaze estimation is complex, as models must account for a continuous variety of human pose and relative camera position. Personal computing applications provide a helpful constraint: the user is typically seated or standing at a position and orientation which stays approximately constant relative to the computer's camera during and between uses. Gaze tracking becomes further simplified when the gaze targets are known; in this way, the system's precision must only be on the scale of the targets themselves, rather than the level of individual pixels. As an additional consideration, for real-time applications such as videoconferencing or video performances, processing steps must be kept minimal to ensure output fast enough for application requirements. In this work, we present a system designed to restore eye contact to the virtual classroom, expanding distance pedagogy capabilities. Stated in a more general frame, we propose a scheme for real-time, calibration-free inference of gaze target for a user-facing camera subject to traditional PC seating and viewing constraints. 

\begin{figure}
    \centering
    \includegraphics[width=75mm]{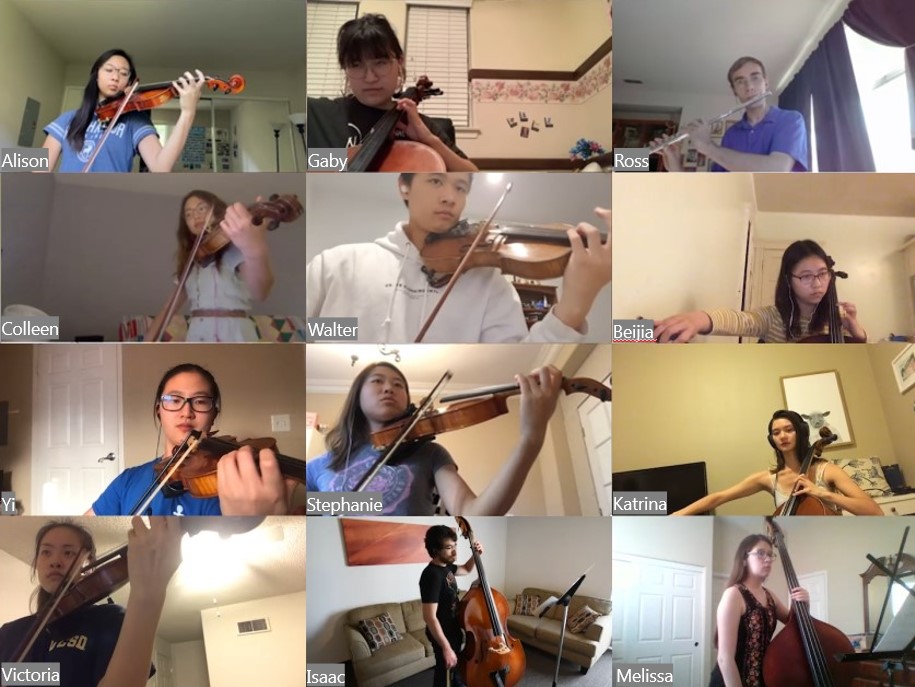}
    \caption{Even though these chamber musicians may look to their concertmaster Alison (upper left) during video rehearsal, her gaze will appear arbitrary as she turns her eyes to a particular player in her view. If she tries to cue by looking at her camera, it will incorrectly appear to all players that they are the intended recipient of her eye contact. From this setup, she is unable to establish a nonverbal communicative connection with another musician.}
    \label{fig:my_label}
\end{figure}

\section{\uppercase{Related Research}}
\begin{figure*}
    \centering
    \includegraphics[width = 160mm]{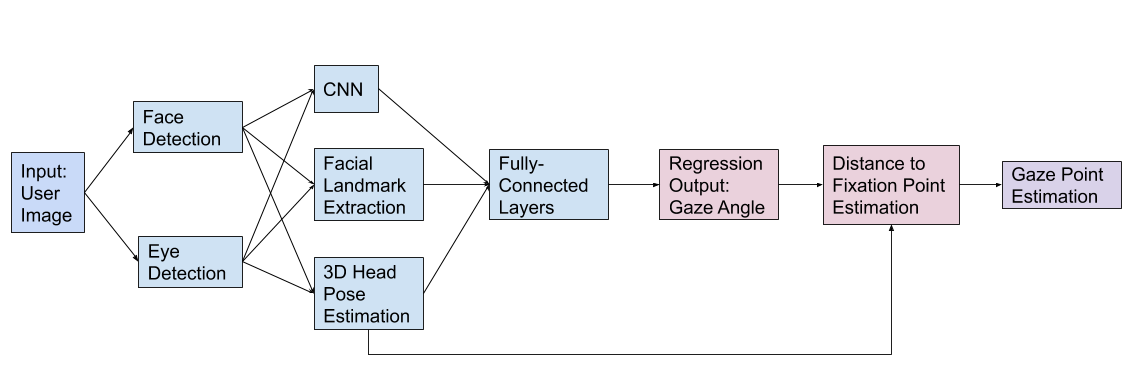}
    \caption{In previous works, a typical pipeline for regression-based gaze estimation methods may involve increased feature extractions to provide enough information to learn the complex output space. Some models discussed in the related works section estimate only up to gaze angle, while others either continue to the gaze point, or bypass the mapping from gaze angle to gaze point and instead learn end-to-end.}
    \label{fig:regpipeline}
\end{figure*}

\subsection{Eye Contact in the Music Classroom}

Many students and educators experienced firsthand the differences between traditional in-person learning environments and synchronous distance learning environments in response to the COVID-19 pandemic. Studies report lower levels of student interaction and engagement, with ``the absence of face-to-face contact in virtual negotiations [leading] students to miss the subtleties
and important visual cues of the non-verbal language." \cite{de2020will}. 
These subtleties are critical to musical education, with past research examining this importance within a variety of musical relationships. 

Within the modality of teacher-ensemble relationships, in review of historical research and their own experiments of conductor eye contact for student ensembles, Byo and Lethco explain the plethora of musical information conveyed via eye contact, its importance in keeping student musicians on task, and its elevated value as musician skill level increases \cite{byo2001student}. In their work, they categorize musical situations and respective usage of eye contact, ranging from use as an attentive tool for group synchronization or an expressive technique for stylized playing. A study by Marchetti and Jensen restates the importance of nonverbal cues as fundamental for both musicians and teachers \cite{davidson2002social}, reframing the value of eye contact and directed gestures within the Belief-Desire-Intention model adopted by intelligent software systems \cite{rao1995bdi}, and pointing to nonverbal communication as a means to provide instruction and correction without interruption \cite{marchetti2010meta}. 

Within the modality of student chamber music, a study of chamber musicians led to categorization of nonverbal behaviors as music regulators (in particular, eye contact, smiles, and body movement for attacks and feedback) ``demonstrated that, during musical performance, eye contact has two important functions: communication between ensemble members and monitoring individual and group performance" \cite{biasutti2013behavioral}. In a more conversational duet relationship, a study of co-performing pianists found eye contact to be heavily used as a means of sharing ideas among comfortable musicians by synchronizing glances at moments of musical importance \cite{williamon2002exploring}. 

Within the modality of improvised music, an examination of jazz students found eye contact to be a vehicle of empathy and cooperation to form creative exchanges, categorizing other modes of communication in improvised music within verbal and nonverbal categories as well as cooperative and collaborative subcategories \cite{seddon2005modes}. Jazz clarinetist Ken Peplowski remarks in \textit{The Process of Improvisation}: ``...we’re constantly giving one another signals. You have to make eye contact, and that’s why we spend so much time facing each other instead of the audience" \cite{peplowski1998process}. Even improvisatory jazz robot Shimon comes with an expressive non-humanoid head for musical social communication, able to communicate by making and breaking eye contact to signal and assist in musical turn-taking \cite{hoffman2010shimon}. 

\subsection{Personal computing gaze classification}

\begin{figure*}
    \centering
    \includegraphics[width=160mm]{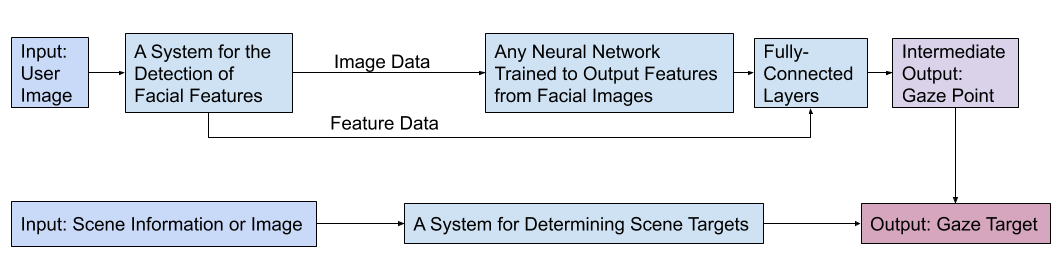}
    \caption{Our pipeline reduces input feature extractions and learns a 2D gaze target coordinate. The simplified model performs with accuracy and speed suitable for videoconferencing applications.}
    \label{fig:mygaze}
\end{figure*}

Vora et al. have demonstrated the effectiveness of using neural networks to classify gaze into abstract ``gaze zones" for use in driving applications and attention problems on the scale of a full surrounding scene \cite{8370646}. Wu et al. use a similar convolutional approach on a much more constrained scene size, limiting gaze zones to the segments spanning a desktop computer monitor \cite{8122270}. In their work, they justify the classification-based model on the basis that the accuracy of regression-based systems cannot reach standards necessary in human-computer interaction (HCI). In their work, they propose a CNN which uses a left or right eye image to perform the classification; to further improve performance, Cha et al. modify the CNN to use both eyes together \cite{8965441}. 

While these approaches are effective in determining gaze to approximately 15 cm by 9 cm zones, this predefined gaze zone approach contains two major shortcomings to identifying the target of a user's gaze. First, multiple objects of interest may lie within one gaze zone, and second, an object of interest may lie on the border spanning two or more gaze zones. Our work addresses these issues by constraining user gaze to a particular target of interest mapped one-per-zone (in the case of the classroom, an individual student musician).

\subsection{Gaze estimation by regression}
In contrast to gaze zone methods, which divide the visual field into discrete regions, regression-based methods better address user attention by opening gaze points to a continuous (rather than discretized) space, but at the expense of taking on a more complex learning task, since the mapping of input image to gaze location becomes many-to-many instead of many-to-few. To address this complexity, many regression models rely on additional input feature extractions from the user image, and regress to estimate the gaze angle, rather than target, to eliminate estimation of an additional uncertainty (distance from eye to object of fixation). Fig. \ref{fig:regpipeline} shows an example pipeline for regression approaches to this problem; though specific works may include or exclude particular features (or stop at gaze angle output), there is still an apparent increase in complexity relative to the end-to-end classification models. 

Zhang et al. and Wang et al. use regression approaches with eye images and 3D head pose as input, estimating 2D gaze angle (yaw and pitch) as output \cite{zhang2015appearance}, \cite{wang2016appearance}. In our work, we generate direct estimation of 2D gaze target coordinates $(x,y)$ relative to the camera position. Zhang et al. contributed a dataset appropriate to our task, MPIIGaze, which contains user images paired with 3D gaze target coordinates; in their work, they use this dataset to again predict gaze angle \cite{zhang2017mpiigaze}. Though we develop our own dataset to meet the assumption of a user seated at a desk, with the optical axis approximately eye level and perpendicular to the user's seated posture, the MPIIGaze dataset could also be used with our method. Sugano et al. use a regression method which does map directly to target coordinates, but their network requires additional visual saliency map input to learn attention features of the associated target image \cite{sugano2012appearance}. 

Closest to our work, Krafka et al. created a mobile-device-based gaze dataset and end-to-end gaze target coordinate regressor \cite{7780608}. Their input features include extracted face, eyes, and face grid, which we reduce to extracted face and 4 bounding box values. They report a performance of 10-15 fps, which falls behind the common videoconferencing standard of 30 fps by a factor of two. In contrast to the previous methods illustrated in Fig. \ref{fig:regpipeline}, our proposed system marries the simplicity of end-to-end, minimal feature classification approaches with the continuous output of regression models to create a fast and accurate solution to gaze target estimation suitable for videoconferencing applications. 

\subsection{Gaze correction}
Some applications may seek to artificially modify the speaker's eye direction to have the appearance of looking directly at the camera, giving the impression of eye contact with the audience. A technology review by Regenbrecht and Langlotz evaluates the benefits of mutual gaze in videoconferencing and presents proposed alterations of workstation setups and hardware to make gaze correction possible. Such schemes include half-silvered mirrors, projectors, overhead cameras, and modified monitors with a camera placed in the center, all of which, while effective, may exceed constraints on simplicity and physical environment for typical remote classroom users \cite{regenbrecht2015mutual}.  

Without modifying hardware, researchers at Intel achieve gaze correction using an encoder-decoder network named ECC-Net \cite{isikdogan2020eye}. The model is effective in its task of modifying a user image such that the eyes are directed toward the camera; it is therefore reasoned that the system intermediately and implicitly learns the gaze direction in the training process. However, their work has two clear points of deviation from our research. First, ECC-Net does not take into account distance from the user to the camera, so though a gaze vector may be possible to infer, the gaze target cannot be directly estimated. Second, while the impression of eye contact is a valuable communication enhancement, our work seeks to communicate to whom the speaker is looking, which is negated if all users assume the speaker to be looking to them. This intermediate step of target determination is critical to communicating directed cues and gestures to individuals. To make this distinction with gaze correction, it becomes necessary to stream a separate video feed of the speaker to each participant with an appropriately corrected (or non-corrected) gaze toward the individual, which is not practical for integration with existing videoconferencing systems that may allow, at best, only for modification of the ego user's singular video feed. 

\section{\uppercase{Methods}}
\subsection{Network Architecture}

Illustrated in Fig. \ref{fig:mygaze}, our proposed system takes a sampled image or video stream as input. The image is first resized and used as input to a face detection system. From this block, we receive extracted facial features to be passed forward in parallel with cropped facial image data. The cropped facial image data is itself passed through a convolutional neural network which extracts further features. The output of the convolutional neural network block is concatenated with the earlier parallel extracted facial features, and passed to a fully-connected block. The output of this fully-connected block corresponds to a regression, mapping the input data to physical locations with respect to the camera origin. Finally, this regression output is fed to a block which determines the user gaze target from a computed list of possible gaze targets. Primarily, our innovations to address gaze detection are in the reduction to lightweight feature requirements and the computation of possible gaze targets (i.e. student musicians in a virtual classroom). In the Implementation Details section, we outline one complete example instance of our general system, which we used in a pilot study with undergraduate students in a remote introductory music course and trained musicians in a remote college orchestra.

\subsection{From Gaze Point to Gaze Target}
Conversion of the regression estimate to gaze target requires scene information. As our work is intended for videoconferencing applications, we have identified two readily available methods to extract scene information:
\begin{itemize}
    \item Process a screencapture image of the user display image to locate video participant faces using a face detection network or algorithm. 
    \item Leverage software-specific system knowledge or API tools to determine video participant cell locations. This may use direct API calls or image processing over the user display image. 

\end{itemize}
From either method, the goal of this step is to generate $N$ coordinate pairs $(x^c_n, y^c_n)$ corresponding to the estimated centroids, measured in centimeters from camera $(0,0)$, of the $N$ target participant video cells  $t_n$. 
Additionally, a threshold $\tau$ is experimentally selected, such that any gaze image whose gaze point is located further than $\tau$ is considered targetless ($t_0$). 
The gaze target $t$ is therefore determined using

\begin{equation}
    n^* = \argminA_n \{(x-x^c_n)^2+(y-y^c_n)^2\}
\end{equation}

\begin{equation}
    t =
    \begin{cases}
    t_{n^*} & (x-x^c_{n^*})^2+(y-y^c_{n^*})^2 \leq \tau^2  \\
    t_0 & \text{otherwise}
    \end{cases}
\end{equation}

A consequence of this method is that the predicted gaze point does not need to be precisely accurate to its target-- the predicted gaze point must simply be closer to the correct target than the next nearest target (and within a reasonable radius).

\subsection{Implementation Details and Training}

\begin{figure*}
    \centering
    \includegraphics[width=160mm]{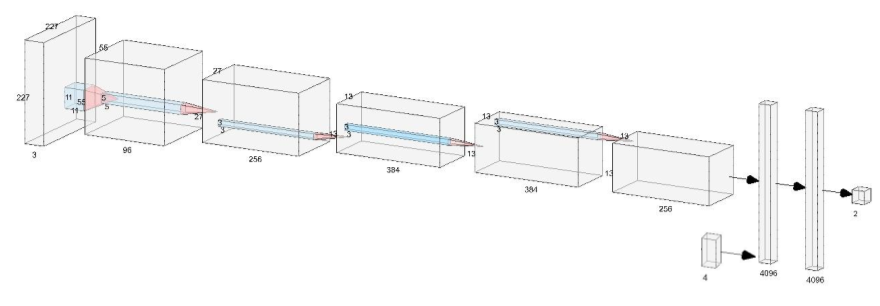}
    \caption{An illustration of one possible facial-image-based convolutional neural network block (the second stage to our system). For our experiments, we use an augmented AlexNet architure. Before the typical fully-connected layers, a 4-vector representing the position and size of the facial bounding box is concatenated to the flattened convolutional output.}
    \label{fig:alexnet}
\end{figure*}

In our experimental instance of the general gaze target detection system, we use as input a stream of images sampled from a USB or integrated webcam, with resolution 1920x1080 pixels. 
Each image is first resized to 300x300 pixels, then fed to a Single-Shot Detector \cite{liu2016ssd} with a ResNet-10 \cite{he2016deep} backbone, pre-trained for face detection. From this network, we receive the upper left corner coordinates, height, and width of the found face's bounding box, which we map back to the original resolution and use to crop the face of the original image. This face-only image is then resized to 227x227 and fed to a modified AlexNet, with the standard convolutional layers remaining the same, but with 4 values (bounding box upper left corner position ($x_b$, $y_b$), bounding box height $h$, and bounding box width $w$) concatenated to the first fully-connected layer. Our architecture is illustrated in Fig. \ref{fig:alexnet}. We propose that the face location and learned bounding box size, combined with the face image, contain enough coarse information about head pose and 3D position for the network to learn the mapping from this input to a gaze point. The 2D output of AlexNet represents the 2D coordinates $(x, y)$, in centimeters, of the gaze point on the plane through the camera perpendicular to its optical axis (i.e. the computer monitor), with the camera located at $(0,0)$. 

For the modified AlexNet stage, our neural network is implemented and trained in PyTorch \cite{baydin2017automatic}. We use a batch size of 32 and Adam optimizer \cite{DBLP:journals/corr/KingmaB14}, using the standard mean squared error loss function. The combined face detection and gaze target model is run in Python using the publicly available, pretrained OpenCV ResNet-10 SSD \cite{opencv_library} for the face detection stage.

\subsubsection{Dataset}

Data was collected from a group of five male and female participants, ages ranging from 20 to 58, with various eye colors. Three of the subjects shared one desktop setup, while the remaining two used individual desktop setups, giving variation in camera position and orientation relative to the subject (in addition to the variation associated with movement in seating position). 91 gaze locations relative to the camera position are defined, spanning a computer screen of dimension 69.84 cm by 39.28 cm. Subjects with smaller screen dimensions were given the subset of gaze locations which fit within their screen. Participants were asked to sit at their desk as they would for typical computer work, and to maintain eye contact with a designated spot on the screen while recording video data. Participants were free to move their head and body to any orientation, as long as eye contact was maintained. 

At this stage in data collection, the participants' videos were stored by gaze location index, with a separate table mapping a gaze location index to its associated displacement from the camera location (measured in centimeters displacement along the horizontal and vertical axes originating from the camera along the screen plane). Each frame of the video for a particular gaze location was then extracted. From each frame, the subject's face was cropped using the aforementioned SSD pretrained on ResNet-10, then resized to 227x227 pixels. The location and dimensions of the facial bounding box are recorded. The dataset is thus comprised of face images, and for each image an associated gaze position relative to the camera and bounding box parameters. 

\begin{figure}
    \centering
    \includegraphics[width=75mm]{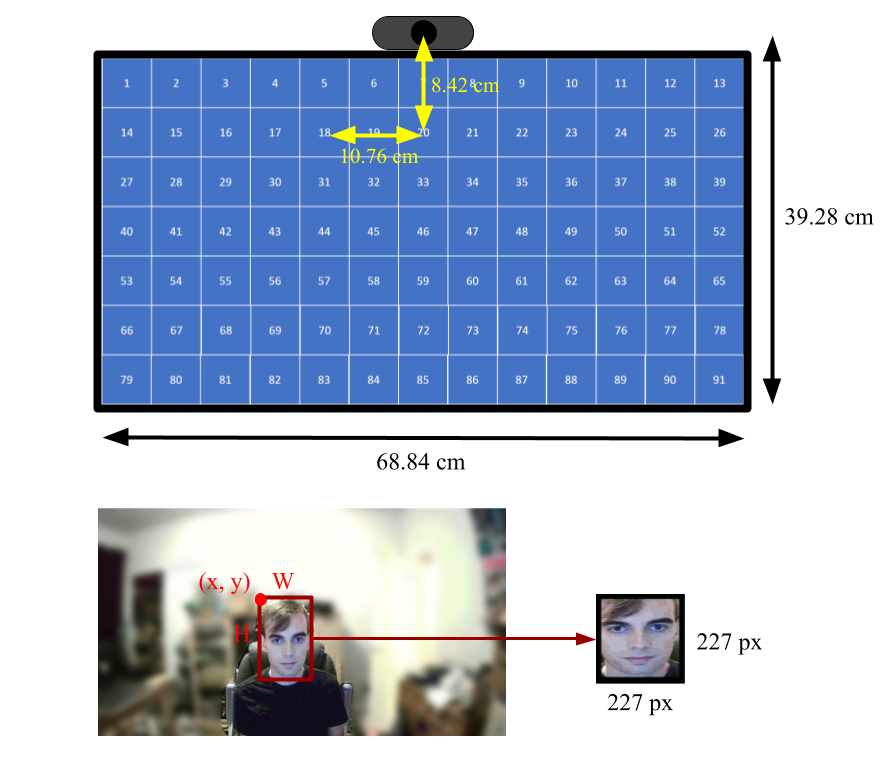} 
    \caption{The data capture setup, including a sample datum. The computer monitor displays different indexed gaze locations for which the subject maintains eye contact. The camera captures a video of the subject for each gaze location. Frames are extracted, with the face then detected and cropped. With each facial image, we store the bounding box parameters (red) as well as the gaze location (yellow).}
    \label{fig:setup}
\end{figure}

A total of 161634 image frames were annotated with one of 91 gaze locations. These images were divided into 129,103 frames for training, 24,398 frames for validation, and 8,133 frames for testing. The data collection setup with an example datum from the training set is shown in Fig. \ref{fig:setup}. 

\subsubsection{Processing Video Layout}

Our system was deployed for experiments over Zoom videoconferencing services \cite{zoom}. Because Zoom offers a set aspect ratio for participant video cells, a consistent background color for non-video-cell pixels, and a fixed location for video participant name within each video cell, we were able to use the second listed method of Section 3.2 (leveraging system knowledge and image processing) to extract gaze targets with the following steps:

\begin{enumerate}
    \item Capture the current videoconference display as an image.
    \item Filter the known background color to make a binary mask.
    \item Label connected components in the binary mask.
    \item Filter out connected components whose aspect ratio is not approximately 16:9.
    \item Find the centroid of each of the remaining connected components

\end{enumerate}

To associate each target with a corresponding index (in the case of our musicians, their name), we crop a predefined region of each discovered video cell, then pass the crop through the publicly available EasyOCR Python package to generate a name label. Depending on the underlying videoconference software employed, such names may be more readily (and accurately) available with a simple API call. 

\subsubsection{Augmenting User Video with Gaze Information}

\begin{figure}
    \centering
    \includegraphics[width=75mm]{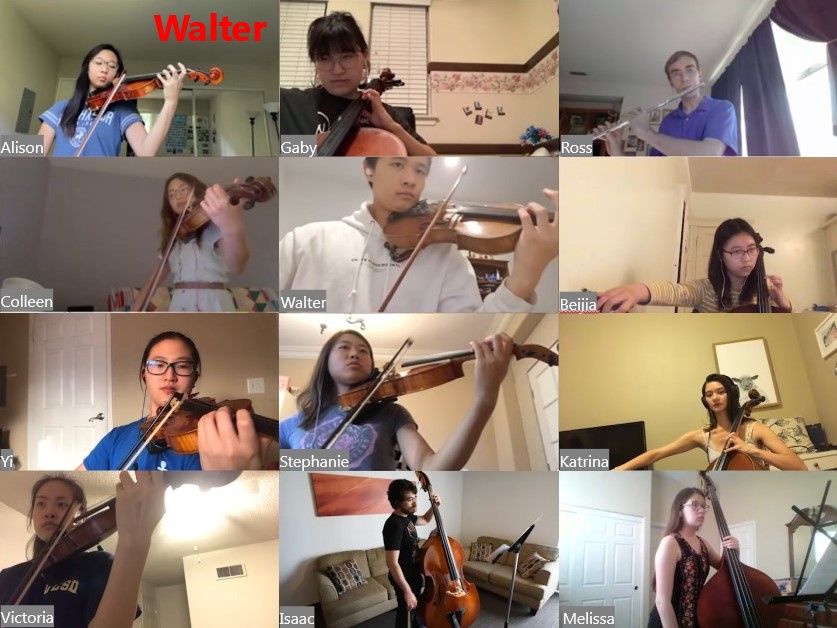}
    \caption{With our system running on her device, concertmaster Alison (upper left) is able to nonverbally communicate to the ensemble that she is looking towards Walter (second row, center). When any students observe Alison's video feed for cues, they will see continuous updates as her gaze changes, establishing nonverbal communication with the ensemble.}
    \label{fig:system_zoom}
\end{figure}

To communicate the user's gaze to call participants, the user's projected video feed is augmented with an additional text overlay. This overlay changes frame-to-frame depending on the estimated gaze target, using the name labels found in the previous section. For our example system, we extract each frame from the camera, add the overlay, then feed each modified frame as a sequence of images to a location on disk. We then run a local server which streams the images from this location, and using the freeware Open Broadcaster Software (OBS) Studio, we generate a virtual camera client to poll this server. The virtual camera is then utilized as the user's selected camera for their virtual classroom, resulting in a displayed gaze as illustrated in Fig. \ref{fig:system_zoom}. 

\section{\uppercase{Experiments}}
\subsection{Computational Performance}
\subsubsection{Inference Time}

Using the Python performance counter timing library, we tested the time for the program to pass a camera image through the face detection and gaze regression model for 1000 trials. The average inference time was approximately 0.03235 seconds, equivalent to 30.9095 frames per second. This satisfies our desired benchmark of 30 fps.  

\subsubsection{Regression Metrics}

The first metric we report is the MSE, i.e. the mean absolute distance from the ground truth gaze location to the predicted gaze location, measured in centimeters relative to the camera. Included with this measure in Table \ref{tab:results} are the absolute distance between truth and prediction in the horizontal (x) direction and the vertical (y) direction. These results show stronger discrimination along the x-axis, motivating the structure of our next video-conference based metrics. 

\begin{table}
\caption{Test Set Error from Gaze Estimate to Truth}
\begin{center}
\begin{tabular}{c c} 
 \hline
 MSE & 1.749 cm \\ 
 \hline
 Average Horizontal Error & 0.678 cm  \\ 
 \hline
 Average Vertical Error & 1.457 cm  \\

\hline
\end{tabular}
\end{center}
\label{tab:results}
\end{table}

The second metric we use is video user hit rate. For each sample, we form a truth target (the nearest video cell to the known gaze point), and a prediction target (the nearest video cell to the estimated gaze point). If these two targets are equal, we classify the sample as a hit, otherwise, a miss. We perform this experiment over two configurations (Zoom Standard Fullscreen Layout and Horizontal Full-Width Layout), over a range of 2 to 8 displayed users. In the Horizontal Full-Width Layout, the video call screen is condensed along the y-axis but kept at full-width along the x-axis, causing all video cells to align in a horizontal row. We maintain this window at the top of the computer screen for consistency. As suggested in Table \ref{tab:results}, because model performance is better at discriminating between gazes along the x-axis, the horizontal layout shows better performance than the fullscreen layout. These hit-rates are provided in Table \ref{tab:results2}. As more users are added to the system, the target points require greater accuracy to form a hit, so performance decreases. Additional variance in performance within a layout class occurs due to the reshaping of the video cell stacking geometry as the number of users change; for example, non-prime numbers of participants lead to a symmetric video cell arrangement, while prime numbers of participants cause an imbalance between rows. When these shapes are well-separated and align around the trained gaze markers, the model performs better, but challenges arise when the video cell layout places borders near trained points. 

\begin{table*}

\caption{Experimental hit-rate of estimated gaze to known targets. }
\begin{center}
\begin{tabular}{c c c c c c c c } 
 \hline
 Number of Displayed Users: & 2 & 3 & 4 & 5 & 6 & 7 & 8 \\ 
 \hline
 Fullscreen Layout Hit Rate: & 0.982 & 0.917 & 0.711 & 0.865 & 0.715 & 0.689 & 0.693 \\ 
 \hline
 Horizontal Layout Hit Rate: & -- & 0.975 & 0.967 & 0.868 & 0.943 & 0.908 & 0.885 \\
 \hline
\end{tabular}

\end{center}

\label{tab:results2}
\end{table*}

\subsection{Classroom Performance}
\subsubsection{Telematic Soundpainting Pilot Setup}
Our system was evaluated among an undergraduate introductory music course, \textit{Sound in Time}. Most students in the survey course are interested in music, but do not have extensive prior musical training. 

Prior to the pilot study, all students in the course were presented with a brief overview lecture of composer Walter Thompson's improvisatory compositional language, \textit{Soundpainting} \cite{thompson2016soundpainting}. In \textit{Soundpainting}, a conductor improvises over a set of predefined gestures to create a musical sketch; students interpret and respond to the gestures using their instruments. For the pilot, student volunteers were assigned at random to one of eight performing groups, with 8-9 students per group (and a final overflow group of 15 musicians), for a total of 72 student subjects. Each of these performing groups was led through a \textit{Soundpainting} sketch by a conductor using the augmented gaze system. 

Throughout each sketch, the performers observed and followed the directions of the conductor, keeping either a conductor-only view or a gallery view on their screens. Within the sketch, students performed from a set of gestures which referred to different groupings of performers, selected from \textit{whole group}, \textit{individuals}, and \textit{live-defined subgroups}.  For whole group play, the conductor would join his hands over his head in an arch, with no specific eye contact necessary. In the case of individuals, the conductor would look at and point to the individual expected to play. In the case of live-defined subgroups, the conductor would look at an individual, assign the individual a symbolic gesture, and repeat for all members of this subgroup. Then, the subgroup's symbol would later be presented, and all members of the subgroup are expected to play. For individual cues and live-defined subgroup formation, the name of the individual being addressed via eye contact is displayed over the conductor, communicating the conductor's intended recipient to the group. 

Four of these group performances were recorded for quantitative analysis, and all students completed a post-performance survey for qualitative analysis. Though the survey was administered by an experimenter outside of the course instructional staff, it is always possible that there may be bias reflected in student responses. Additionally, students unfamiliar with \textit{Soundpainting} may provide responses which address a combination of factors from both the augmented gaze system and their level of comfort with the \textit{Soundpainting} language. This same unfamiliarity may have also affected students' ability to respond to gestures for quantitative analysis; a student who forgets a gesture's meaning may freeze instead of playing, even if they recognize it is their turn.  
\subsubsection{Quantitative Measures}

During the sketches of the four recorded performing groups, a total of 43 cues were delivered from the conductor to the musicians. For each cue and for each musician, a cue-musician interaction was recorded from the following:
\begin{itemize}
    \item Hit (musician plays when cued)
    \item Miss (musician does not play when cued)
    \item False Play (musician plays when not cued)
    \item Correct Reject (musician does not play, and musician was not cued)
\end{itemize}

In cases where the student was out of frame and thus unobservable, data points associated with the student are excluded. In total, 114 cue-musician interactions were observed. Binary classification performance statistics are summarized in Table \ref{tab:results_cues}. Hit Rate describes the frequency in which students play when cued, while Miss Rate describes the frequency in which students fail to play when cued. Precision describes the proportion of times a musician was playing from a cue, out of all instances that musicians were playing (that is, if precision is high, we would expect that any time a musician is playing they had been cued). Accuracy describes the likelihood that a student's action (whether playing or silent) was the correct action. While there is no experimental benchmark, our statistics suggest that the musicians were responsive to cues given to individuals and groups through nonverbal communication during their performances.   

In addition to recorded performance metrics, student responses to the post-performance survey indicated that 87\% of respondents felt they established a communicative relationship with the conductor (whose gaze was communicated), while only 42\% felt they established a communicative relationship with other musicians (who they could interact with sonically, but without eye contact). Reflecting on previously cited works discussing the importance of eye contact in improvised and ensemble music, our experimental results suggest that having eye contact as a nonverbal cue significantly improves the establishment of a communicative relationship during performance in distance learning musical environments. 

\begin{table}

\caption{Student-Cue Response Statistics (n = 114)}
\begin{center}
\begin{tabular}{c c} 
 \hline
 Hit Rate & 0.851 \\ 
 \hline
 Miss Rate & 0.149  \\
 \hline
 Precision & 0.990  \\ 
 \hline
Accuracy & 0.952  \\
 \hline
\end{tabular}

\end{center}

\label{tab:results_cues}
\end{table}

\subsubsection{Qualitative Results}

Overall, students were able to successfully perform \textit{Soundpainting} sketches over Zoom when conducted with an augmented gaze system. In addition to undergraduate students in the pilot study, trained student musicians of a college orchestra gathered on Zoom to give a similar performance. The group of trained musicians responded positively to the performance, commenting that this was the first time they felt they were partaking in a live ensemble performance since the beginning of their distance learning, since previous attempts at constrained live performance tended to be meditative rather than interactive. Student feedback towards performing with the augmented gaze system include the following remarks: 
\begin{itemize}
    \item ``Able to effectively understand who [the conductor] was looking at"
    \item ``Easy to follow because the instructions were visible"
    \item ``It was as good as it would get on Zoom, in my opinion. We were able to see where [conductor] was looking (super cool!) and I think that allowed us to communicate better."
    \item ``...Easy for us to understand the groupings with the name projection..."
    \item ``I was able to communicate and  connect with the conductor, because it was essentially like a conversation, where he says something and I answer."
    \item ``All I had to do was pay attention and follow"
\end{itemize}
To an ensemble musician, these remarks suggest a promising possibility of interactive musicianship previously lost in the virtual classroom by restoring channels of nonverbal communication to the distanced ensemble environment. 

\section{FUTURE WORK}

To improve technical performance of the gaze target system in both precision and cross-user accuracy, a logical first step is the extension of the training dataset. Collecting data on a diverse set of human subjects and over a greater number of gaze targets can provide greater variance to be learned by the model, more closely resembling situations encountered in deployment. Augmenting our miniature dataset with the aforementioned MPIIGaze dataset may also improve model performance and provide an opportunity to compare performance to other models. 

In addition to increasing the number of subjects in the dataset, there is also room for research into optimal ways to handle pose diversity. Musicians perform with different spacing, seating, and head position depending on their instrument, and having a gaze system which adapts according to observed pose for increased accuracy is important for use among all musicians. 

To improve inference time of the system, future research should include experimentation with multiple neural network architectures. In this work, we use an AlexNet base, but a ResNet or MobileNet architecture may also learn the correct patterns while reducing inference time. Other extracted features may also prove useful to fast and accurate inference.

Further experiments that would provide information about system performance in its intended environment would include testing on computer desk physical setups which are not part of the training dataset to capture a diversity of viewing distances and angles, and testing the system on human subjects who are not part of the training dataset to measure the system's ability to generalize to new faces and poses.

From an application standpoint, there is a need to expand communication channels from conductor-to-musician (or speaker-to-participant) to musician-to-musician (or participant-to-participant). While the current system was designed around a hierarchical classroom structure, peer-to-peer interaction also provides value, and adding this capability will require  changes in the indication design to avoid overwhelming a student with too many sources of information. Practically speaking, development of a plugin which integrates with existing video conferencing software and works cross-platform is another area for further development. 

In addition to musical cooperation, another application of the system lies in gaze detection to manage student attention. Such a system could provide a means for a student to elect to keep their camera off to viewers, while still broadcasting their attention to the presenter. Such a capability would allow teachers to monitor student attention even in situations when students wish to retain video privacy among their peers. 

Finally, ethical considerations of the proposed system must be made a priority before deployment in any classroom. Two particular ethical concerns which should be thoroughly researched are the ability of the system to perform correctly for all users (regardless of difference in physical appearance, skin color, or ability), and provisions to protect privacy on such a system, including the readily available option for a user to turn off the gaze tracking system at any time. 

\section{CONCLUDING REMARKS}

In this paper, we introduced a fast and accurate system by which the object of a user's gaze on a computer screen can be determined. Instantiating one possible version of the system, we demonstrated the capability of an extended AlexNet architecture to estimate the gaze target of a human subject seated in front of a computer using a single webcam. Possible applications of this system include enhancements to videoconferencing technology, enabling communication channels which are otherwise removed from telematic conversation and interaction. By using publicly available Python libraries, we read video user names and cell locations directly from the computer screen, creating a system which both recognizes and communicates the subject of a person's gaze. 

This system was then shown to be capable of restoring nonverbal communication to an environment which relies heavily on such a channel: the music classroom. Through telematic performances of Walter Thompson's \textit{Soundpainting}, student musicians were able to respond to conducted cues, and experienced a greater sense of communication and connection with the restoration of eye contact to their virtual music environment.

\section*{\uppercase{Acknowledgements}}

The authors would like to thank students and teaching staff from the \textit{Sound in Time} introductory music course at the University of California, San Diego; musicians from the UCSD Symphonic Student Association; and students of the Southwestern College Orchestra under the direction of Dr. Matt Kline. 

This research was supported by the UCOP Innovative Learning Technology Initiative Grant \cite{ucop}. 

\bibliographystyle{apalike}
{\small
\bibliography{example}}

\end{document}